\documentstyle[jkas41]{article} 

\runningauthor{KYEONG ET AL.} \runningtitle{OPEN CLUSTER NGC 1193} 
\beginpage{1} 
\endpage{9}

\def\arcdeg{\hbox{$^\circ$}} 
\def\arcmin{\hbox{$^\prime$}} 
\def\arcsec{\hbox{$^{\prime\prime}$}}

\begin{document} 

\title{$UBVI$ CCD Photometry of the Old Open Cluster NGC 1193} 

\author{ 
{\normalsize\textbf{\textsc{J}}\scriptsize\textbf{\textsc{AEMANN}}} 
{\normalsize\textbf{\textsc{K}}}\scriptsize\textbf{\textsc{YEONG}}$^1$ 
\textbf{\textsc{,}} 
{\normalsize\textbf{\textsc{S}}}\scriptsize\textbf{\textsc{ANG}} 
{\normalsize\textbf{\textsc{C}}}\scriptsize\textbf{\textsc{HUL}} 
{\normalsize\textbf{\textsc{K}}}\scriptsize\textbf{\textsc{IM}}$^{1,2}$ 
\textbf{\textsc{,}} 
{\normalsize\textbf{\textsc{D}}\scriptsize\textbf{\textsc{AVID}}} 
{\normalsize\textbf{\textsc{H}}}\scriptsize\textbf{\textsc{IRIART}}$^3$ 
\textbf{\textsc{AND}} 
{\normalsize\textbf{\textsc{E}}}\scriptsize\textbf{\textsc{ON}}$-${\normalsize\textbf{\textsc{C}}}\scriptsize\textbf{\textsc{HANG}} 
{\normalsize\textbf{\textsc{S}}}\scriptsize\textbf{\textsc{UNG}}$^1$ 
} 

\offprints{J. Kyeong} 

\address{$^1$Korea Astronomy \& Space Science Institute, Daejeon 305-348, Korea} 
\address{\it E-mail: jman, sckim, ecsung@kasi.re.kr} 
\address{$^2$Department of Astronomy and Astrophysics, University of Toronto, Toronto, ON M5S 3H4, Canada}
\address{\it E-mail: sckim@astro.utoronto.ca} 
\address{$^3$Instituto de Astronom\'{\i}a, Universidad Nacional Aut\'onoma de
  M\'exico, Campus Ensenada, Apdo. Postal 877, 22800 Ensenada, B.C., Mexico} 
\address{\it E-mail: hiriart@astrosen.unam.mx} 

\vskip 3mm 

\address{\normalsize{\it (Received ???. ??, 2008; Accepted ???. ??, 2008)}} 

\abstract{ 
We present $UBVI$ photometry of the old open cluster NGC 1193. 
Color-magnitude diagrams (CMDs) of this cluster show 
a well defined main sequence and a sparse red giant branch. 
For the inner region of $r<50\arcsec$, three blue straggler candidates are newly found 
in addition to the objects Kaluzny (1988) already found.  
The color-color diagrams show that the reddening value toward NGC 1193 is 
$E(B-V) =0.19 \pm 0.04$. 
From the ultraviolet excess measurement, 
we derived the metallicity to be [Fe/H]$=-0.45 \pm 0.12$. 
A distance modulus of $(m-M)_0 =13.3 \pm 0.15$ 
is obtained from zero age main sequence fitting with 
the empirically calibrated Hyades isochrone of Pinsonneault et al. (2004). 
CMD comparison with the Padova isochrones by Bertelli et al. (1994) 
gives an age of log $t =9.7 \pm 0.1$. 
} 


\maketitle 

\section{INTRODUCTION} 

Since the open clusters are composed of stars with almost the same age and metallicity, 
they can be used in testing the theoretical models for stellar evolution. 
The relatively close open clusters can be good targets for even small telescopes 
to get a good enough quality photometric data (e.g.: Kyeong, Byun, \& Sung 2001; 
Kim \& Sung 2003). 
In particular, the old open clusters are very useful in testing the stellar isochrones 
because they show clear evolutionary sequences. 

There have been a few observational studies on NGC 1193, which has a condensed central 
region.  One is the $BV$ photometry of Kaluzny (1988) which is used to derive 
the basic parameters.  The other is Friel \& Janes (1993)'s spectroscopic study 
of several stars of NGC 1193 to determine its metallicity. 

In this paper, we describe the $UBVI$ CCD observations of the old open cluster NGC 1193, which 
is located at $\alpha$=3$^h$ 05$^m$, $\delta$=$+$44$^\circ$ 23$'$ (J2000). 
Although the bright star in the western 
side hamper the wide field observation, we present accurate $UBVI$ photometry for this 
cluster and derive the basic parameters such as reddening, metallicity, distance modulus 
and age from the photometric data.  The robust determination of reddening 
is significant for the estimation of distance, and subsequently age and metallicity. 
So, we estimate the reddening through two-color diagram method ($U-B, B-V$) 
and other cluster parameters are then derived by various conventional methods. 

Section II describes the observations and data reduction processes. 
The comparison with the previous photometric data is presented in Section III. 
Color-magnitude diagrams (CMDs) and blue straggler candidates are explained in Section IV. 
Cluster parameters such as the reddening, metallicity, distance modulus and 
age are derived in Section V.  Section VI summarizes our results. 

% -------------------------------------------------- TABLE 1 START 
\begin{table*} 
\begin{center} 
{\bf Table 1.}~~Journal of Observations for NGC 1193 \\ 
\begin{tabular}{ccccc} 
\hline \hline 
\noalign{\smallskip} 
Date &UT(Start) & Filter & ${\rm T_{exp}}$(sec) &Airmass \\ 
\noalign{\smallskip} 
\hline 
\noalign{\smallskip} 
& 11:06 &$U$ &600 $\times$ 3 &1.06 \\ 
2006 August 30 & 10:50 &$B$ &300 $\times$ 3 &1.07 \\ 
& 11:38 &$V$ &200 $\times$ 3 &1.04 \\ 
& 10:37 &$I$ &60 $\times$ 3 &1.09 \\ 
\noalign{\smallskip} 
\hline 
\noalign{\smallskip} 
\end{tabular} 
\end{center} 
\end{table*} 
%-------------------------------------------------- TABLE 1 END 

%-------------------------------------------------- FIG 1 START 
\begin{figure*}%[p] 
\vskip 5mm 
\epsfxsize=8.9cm 
\epsfysize=8.9cm 
\centerline{\epsffile{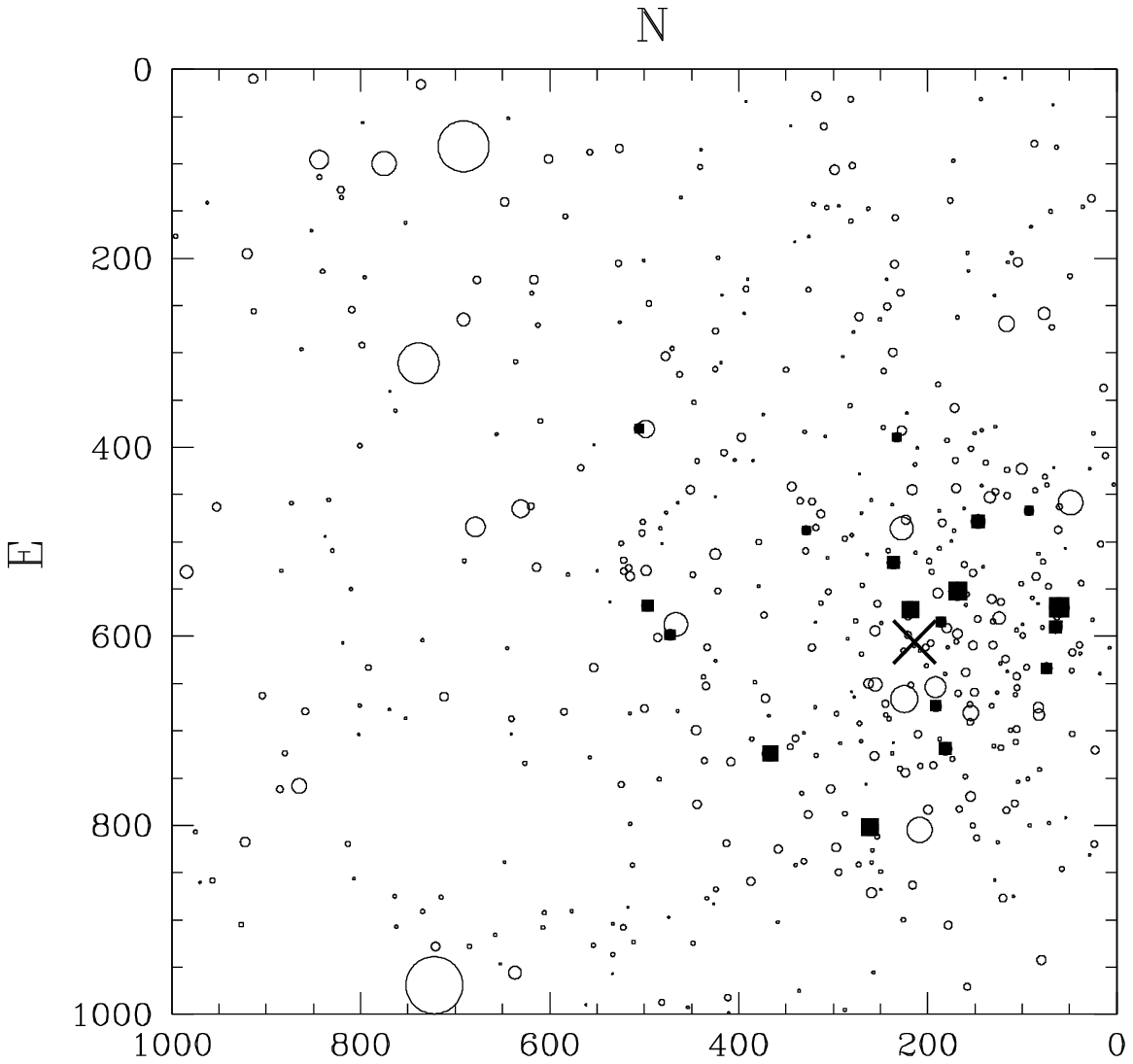}} 
{\small {\bf ~~~Fig. 1.}---Synthetic field (4$\farcm$3$\times$4$\farcm$3) of 
NGC 1193 as observed in $V$ frame.  North is at the top and east is to the left. 
All the detected stars are plotted.  The X- and Y-axes are in pixels and 1 pixel corresponds to $0\farcs25$. 
Circle and square sizes are proportional to the brightnesses. 
The large cross means the cluster center determined by Kaluzny (1988). 
The filled squares represent the blue straggler candidates (See the text), 
while their sizes are also proportional to their brightnesses. 
} 
\end{figure*} 
%-------------------------------------------------- FIG 1 END 

\section{OBSERVATIONS AND DATA REDUCTION} 

$UBVI$ CCD images of the open cluster NGC 1193 were obtained on a photometric night, 
2006 August 30, using the 1.5m telescope (f/13.5) and SITe 1K CCD at the Observatorio 
Astron\'omico Nacional in the Siera San Pedro M\'atir(OAN-SPM) in Baja California, 
Mexico. 
The pixel scale is 0$\farcs$25/pixel with a total field of view 
of 4$\farcm$3$\times$4$\farcm$3.  The gain and readout noise of the CCD camera are 1.2 $e^-$/ADU and 
7.8 e$^-$, respectively.  The median seeing was around 1$''$. 
Specially, in order to get the long exposures, we excluded the bright 
star ($V=11.23$ mag, $K=6.1$ mag) on the west of NGC 1193. 
The journal of the observation of NGC 1193 is given in Table 1. 

%-------------------------------------------------- FIG 2 START 
\begin{figure*}%[p] 
\vskip 5mm 
\epsfxsize=8.9cm 
\epsfysize=8.9cm 
\centerline{\epsffile{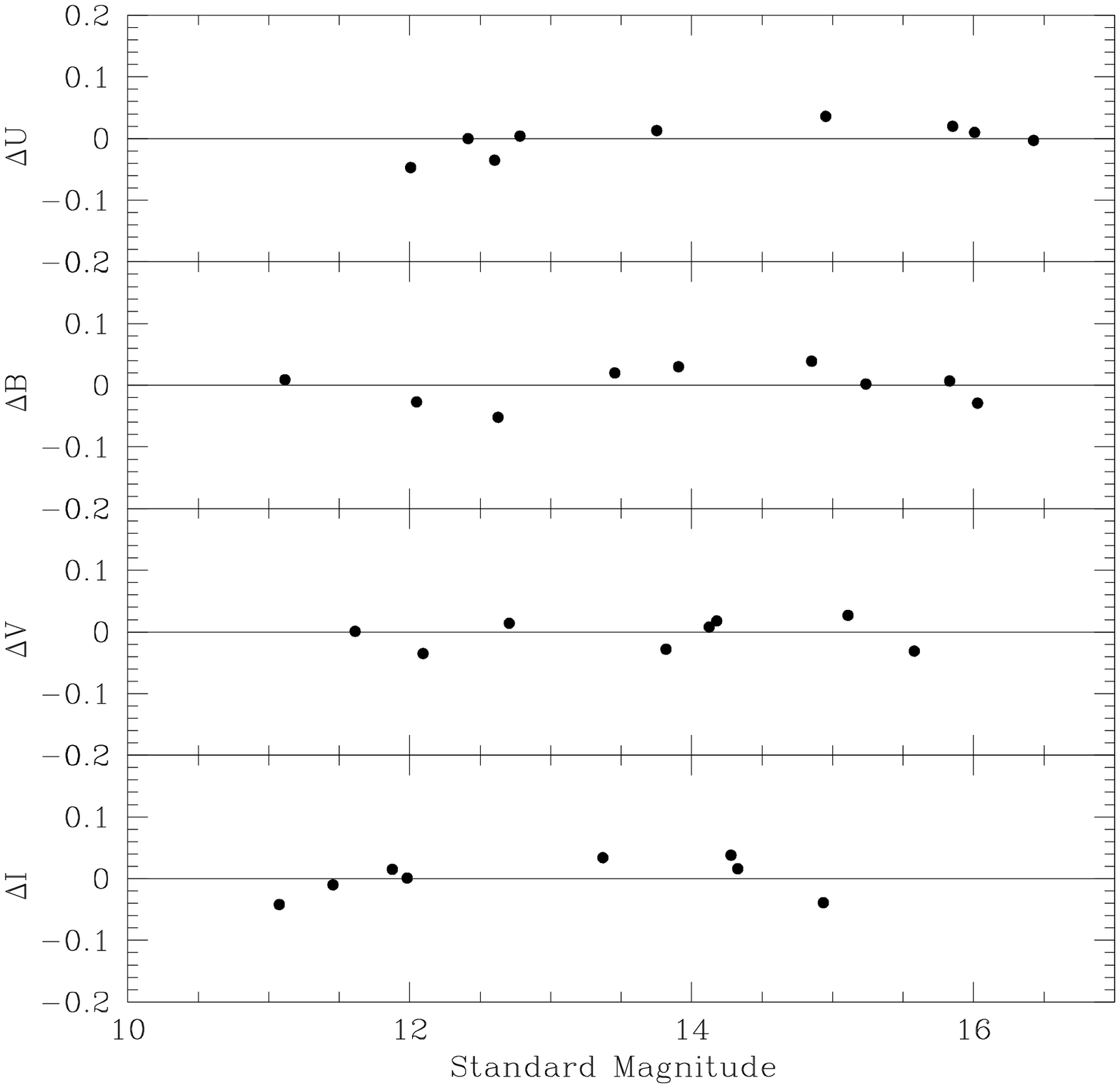}} 
{\small {\bf ~~~Fig. 2.}---Residuals between standard and transformed magnitudes 
of standard stars, plotted against standard magnitudes. $\Delta$ means standard magnitude
minus transformed magnitude.
} 
\end{figure*} 
%-------------------------------------------------- FIG 2 END 

% -------------------------------------------------- TABLE 2 START 
\begin{table*} 
\begin{center} 
{\bf Table 2.}~~Derived standardization coefficients and its errors \\
\begin{tabular}{ccccc} 
\hline \hline
\noalign{\smallskip} 
Filter &$k_1$ & C$_1$ & Zero point \\
\noalign{\smallskip} 
\hline 
\noalign{\smallskip} 
$U$ & 0.664$\pm$0.028 &$+$0.035$\pm$0.014  &$-$2.708$\pm$0.009 \\ 
$B$ & 0.313$\pm$0.005 &$+$0.094$\pm$0.024  &$-$1.579$\pm$0.019 \\ 
$V$ & 0.283$\pm$0.021 &$-$0.008$\pm$0.019  &$-$1.075$\pm$0.015 \\ 
$I$ & 0.162$\pm$0.030 &$+$0.039$\pm$0.020  &$-$1.271$\pm$0.019 \\  
\noalign{\smallskip}  
\hline 
\noalign{\smallskip}  
\end{tabular}
\end{center} 
\end{table*} 
%-------------------------------------------------- TABLE 2 END

Standard procedures using the IRAF\footnote{IRAF (Image Reduction and Analysis Facility) 
is distributed by the National 
Optical Astronomy Observatory, which are operated by the Association of 
Universities for Research in Astronomy, Inc., under cooperative agreement 
with the National Science Foundation.} CCDRED package were followed. 
Object CCD 
frames were flattened using twilight sky flats after overscan correction and bias 
subtraction. 
For each filter, the three frames were median-combined to remove the cosmic 
rays and to get the high signal-to-noise ratio. 
Instrumental magnitudes of stars were 
obtained by the point-spread function (PSF) fitting packages of DAOPHOT II and 
ALLSTAR (Stetson 1990) 
and stars of different frames were matched by DAOMATCH/DAOMASTER routines (Stetson 1992). 
For each frame, a representative model PSF was constructed by using a number of 
unsaturated and relatively isolated stars. 
Due to the small field of view of the CCD, it was not necessary 
to allow the shape of the model PSF to vary quadratically with position in each frame. 
Finally, aperture corrections were made using the DAOGROW program (Stetson 1990), 
for which we used the same stars used in the PSF construction. 

The observed CCD field of the cluster is shown in Fig. 1. 
Each star is represented by an open circle, whose size is proportional to the brightnesses. 
We use the value (X$_C$, Y$_C$) =(605, 214) determined by Kaluzny (1988) as the center of the cluster. 
Of course their coordinates are converted by the transformation equation 
between two coordinate systems. 

Additionally, we observed 14 standard stars of four Landolt (1992) regions 
(PG2113, SA112, SA92$\times$2). 
The color range of these standard stars is $-0.2<(B-V)<1.4$ 
which fully covers the color range of the stars in NGC 1193. 
Standard stars were observed at airmasses between 1.1 and 2.4. 

With these standard star observations, the transformation equations between 
standard and instrumental magnitudes are as follows: 

\begin{eqnarray*} 
U & = & u - k_{1U} ~X_U + C_U ~(U-B) + Z_U  \\
B & = & b - k_{1B} ~X_B + C_B ~(B-V) + Z_B  \\
V & = & v - k_{1V} ~X_V + C_V ~(B-V) + Z_V  \\
I & = & i - k_{1I} ~X_I + C_I ~(V-I) + Z_I  \\
\end{eqnarray*} 
\noindent 
where small letters represent instrumental magnitudes and the capital 
letters indicate standard magnitudes. X is the airmass of each filter. 
All coefficients and its errors are derived from our own standard 
star observations are also given in Table 2. 

Fig. 2 shows the standardization residuals for $UBVI$ filters. 
We inspected the particular trends like UT term but didn't find any trends in the residuals. 
In particular, there was no dependency on the secondary extinction coefficient in $U$ and $B$
band, so only the primary extinction coefficients are used in the standardization.

%-------------------------------------------------- FIG 3 START 
\begin{figure*}%[p] 
\vskip 5mm 
\epsfysize=7.0cm 
\centerline{\epsffile{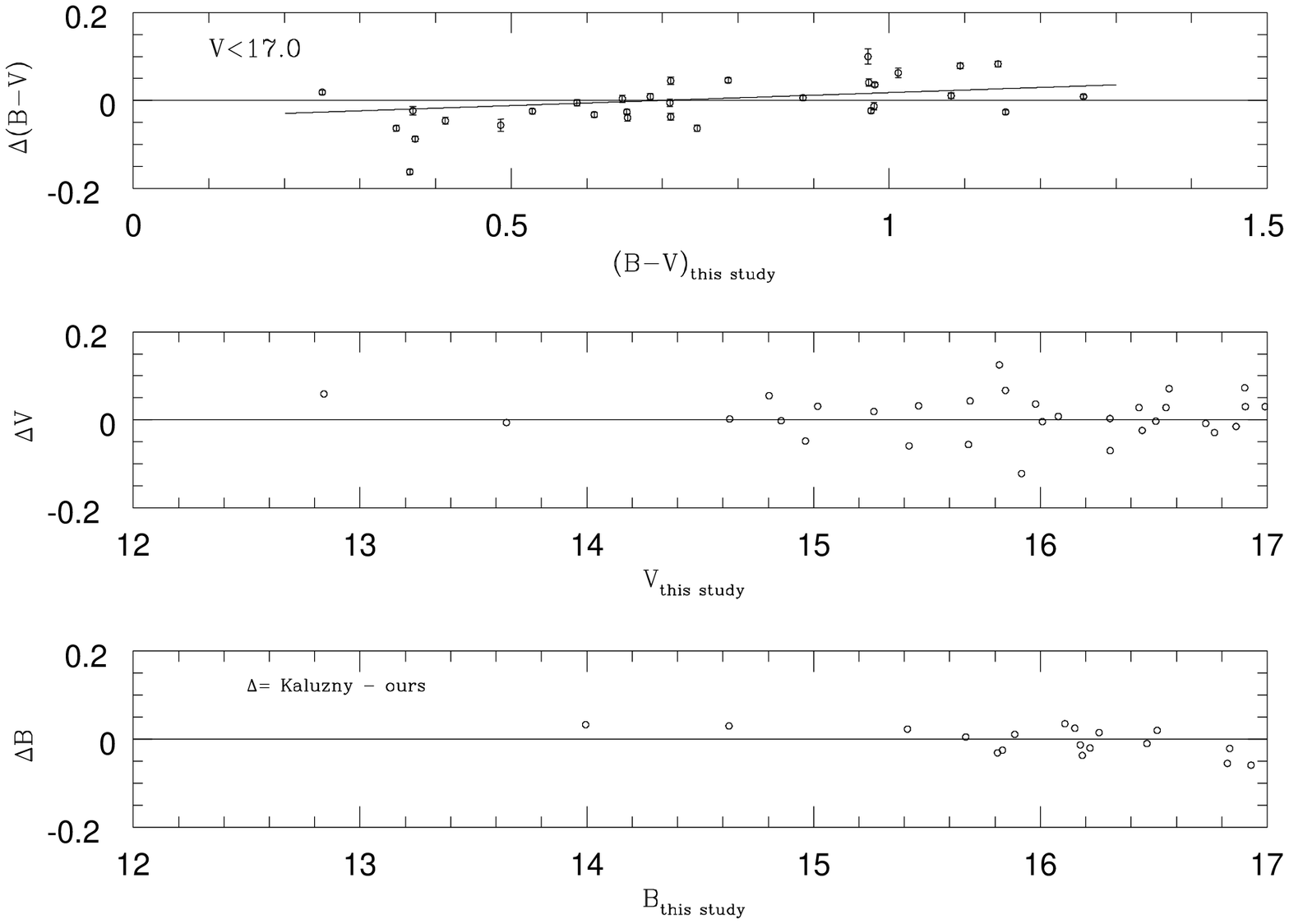}} 
{\small {\bf ~~~Fig. 3.}---Comparison of our photometry with that of Kaluzny (1988), 
where $\Delta$ means Kaluzny minus this study.  For the case of $\Delta(B-V)$ ($upper~panel$), 
a linear fitting is overlaid to show the trend clearly. 
} 
\end{figure*} 
%-------------------------------------------------- FIG 3 END 

\section{COMPARISON WITH THE PREVIOUS PHOTOMETRY} 

Fig. 3 shows the comparison of the photometries between this study and 
the $BV$ photometry Kaluzny (1988). 
For the comparison, we have transformed the X- and Y-coordinates 
between ours and Kaluzny's using the position of several bright stars. 

Fig. 3 shows that the differences of $B$ and $V$ magnitudes of stars with $V<17$ 
between the two datasets are negligible within the standardization residuals. 
On the other hand, for the case of $\Delta(B-V)$ ($upper~panel$), 
a trend is shown.  In particular, there are two maximum differences of 
$\Delta(B-V)=-0.02$ at the blue color ($B-V=0.30$) and 
$\Delta(B-V)=+0.040$ at the red color ($B-V=1.3$). 
This trend means that these color differences make 
the bluer (than the turn-off) stars bluer and make 
the red giant stars redder in Kaluzny (1988)'s CMDs. 
Therefore, the age of NGC 1193 derived by Kaluzny (1988) 
using the morphological age index (MAI) method or 
isochrone fitting method appears to be somewhat young. 

The photometric data of Kaluzny (1988) are slightly undersampled 
(with the relatively large pixel scale of $0\farcs86$/px compared to the seeing of 
$\sim 1\farcs5$), 
and the systematic errors of their data are as large as 0.06 mag 
which is resulted from the uncertainties in their standard transformations and 
those in their aperture corrections. The source of this trend in $\Delta(B-V)$ 
is not clear. 

%-------------------------------------------------- FIG 4 START 
\begin{figure*}%[p] 
\vskip 5mm 
\epsfysize=7.0cm 
\centerline{\epsffile{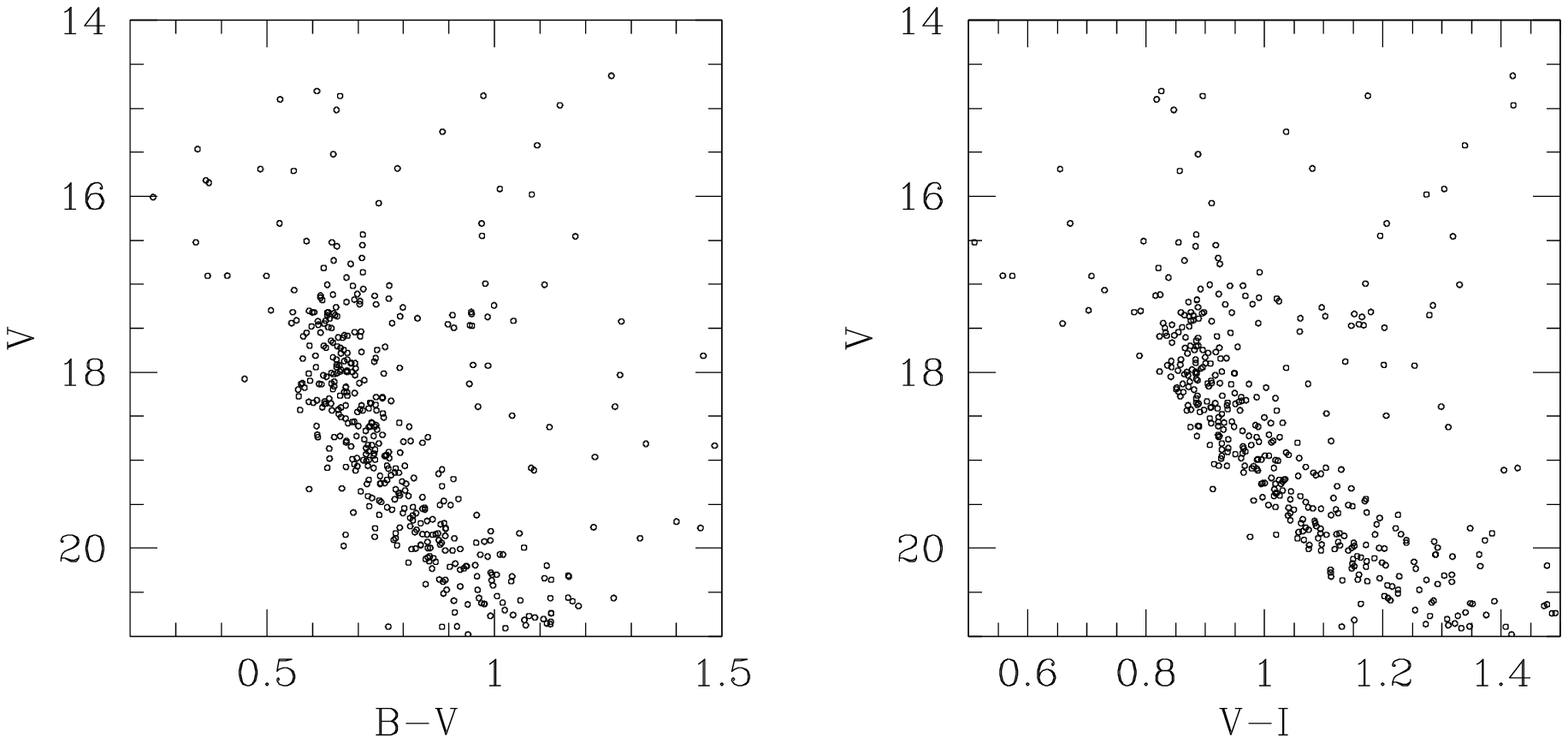}} 
{\small {\bf ~~~Fig. 4.}---Color-magnitude diagrams of NGC 1193. 
} 
\end{figure*} 
%-------------------------------------------------- FIG 4 END 

%-------------------------------------------------- FIG 5 START 
\begin{figure*}%[p] 
\vskip 5mm 
\epsfxsize=8.9cm 
\epsfysize=8.9cm 
\centerline{\epsffile{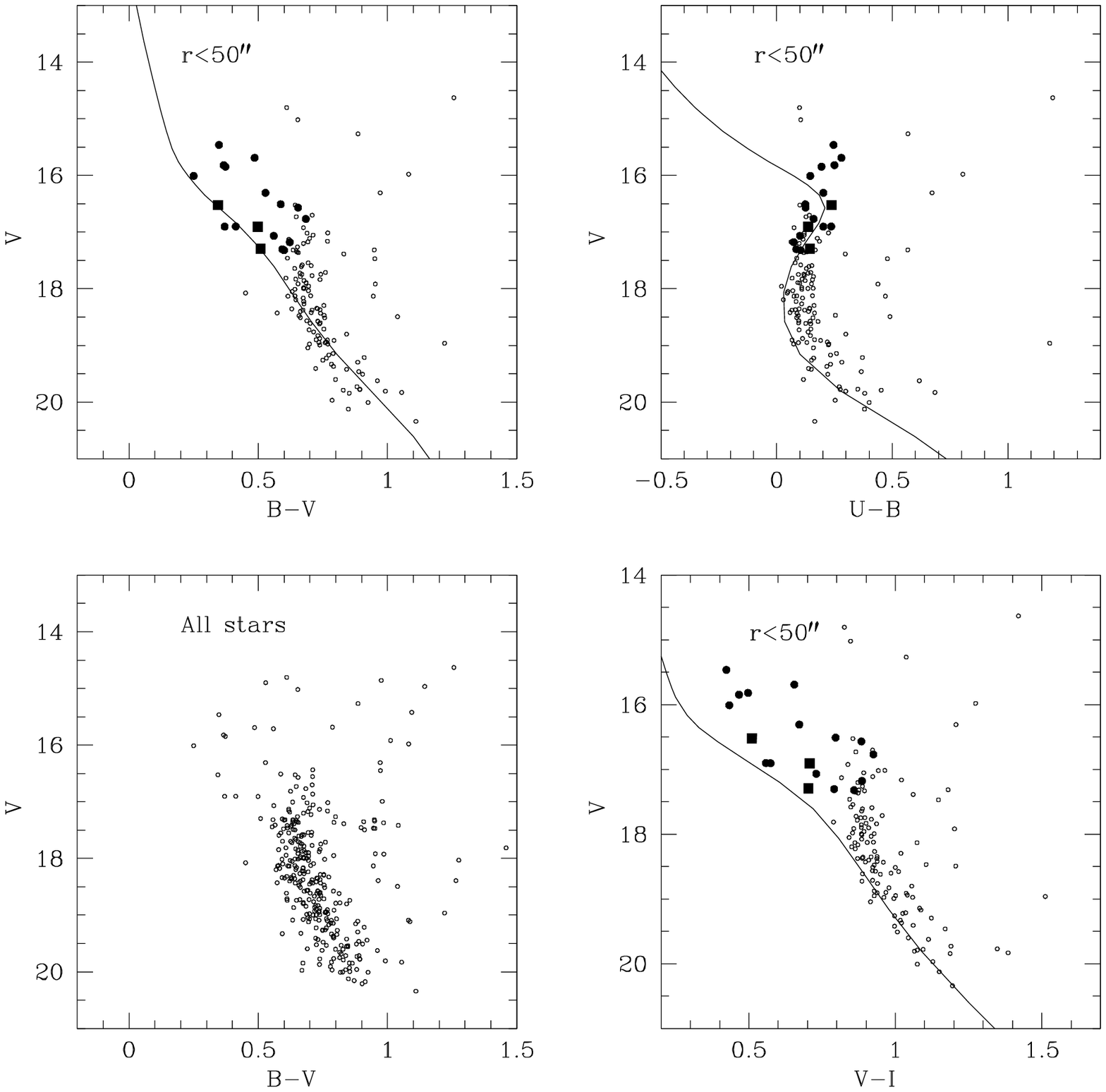}} 
{\small {\bf ~~~Fig. 5.}---Blue straggler candidates of NGC 1193 are denoted 
as larger filled symbols on the optical-band CMDs. 
Filled squares are three blue straggler candidates newly found in this study 
and filled circles are the candidates already known in the literatures, 
while small open circles are all other stars. 
Solid lines denote the ZAMS line by Bertelli et al.(1994) 
with [Fe/H]$=-0.61$, and log $t=7.8$. 
} 
\end{figure*} 
%-------------------------------------------------- FIG 5 END 

\section{COLOR MAGNITUDE DIAGRAMS AND BLUE STRAGGLER CANDIDATES} 

331 stars are detected in all the $UBVI$ optical bands. 
This small number of stars is mainly due to the scarcity of detected stars 
in the $U$ filter. In Fig. 4 we present CMDs in various passbands.
But the CMDs of NGC 1193 show the distinguishable features usually seen in 
the CMDs of old open clusters (e.g., NGC 2243, NGC 188, etc): 
well defined main sequence (MS), sub giant branch and red giant branch. 

Ahumada \& Lapasset (1995) define the blue straggler region as 
$0.0 < (B-V)_0 < 0.8$ and $1.0 < M_V < 5.0$, 
located above the zero age main sequence (ZAMS) line. 
They pointed out that 
the blue stragglers seems to be concentrated in the central region of the clusters 
and the degree of concentration is stronger in the case of old open clusters. 
Adopting the distance to NGC 1193 derived in this study (see \S V), $(m-M)_0=13.32$, 
$M_V$ range for the blue stragglers in NGC 1193 becomes $14.3 - 18.3$. 
We identified the 15 blue straggler candidates already known 
in literatures\footnote{WEBDA (http://www.univie.ac.at/webda/) report 
16 blue straggler candidates in NGC 1193, 
while one is outside of the field of view of our CCD frame.} 
including the five stars discovered by Kaluzny (1988) 
and identified three new objects by applying the Kaluzny (1988) and Ahumada \& Lapasset
(1995)'s criterion.  Fig. 5 shows these blue straggler candidates in CMDs. 
The $UBVI$ magnitudes of these blue stragglers are given in Table 3. 

%-------------------------------------------------- FIG 6 START 
\begin{figure}%[p] 
\vskip 5mm 
\epsfysize=7cm 
\centerline{\epsffile{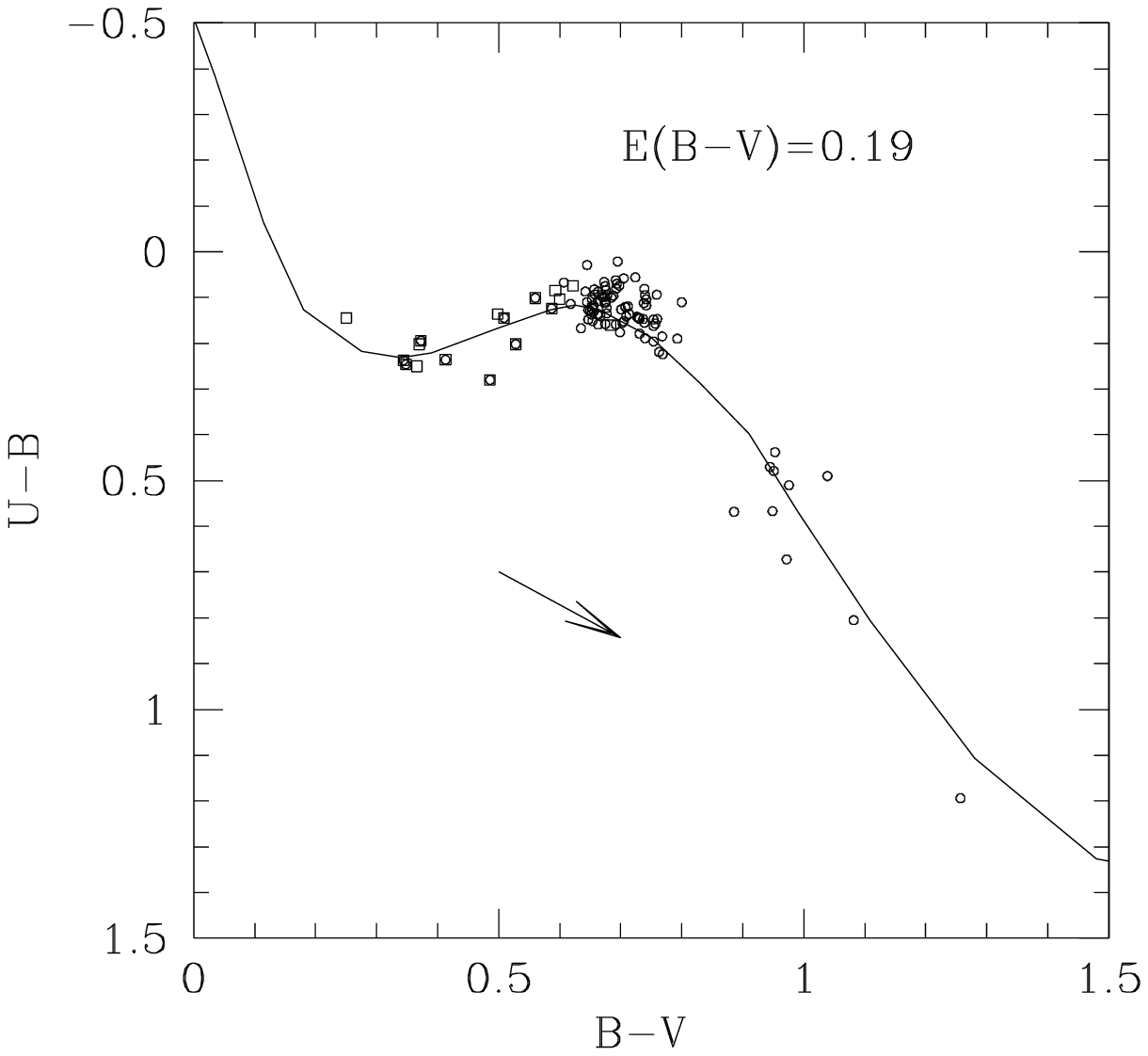}} 
{\small {\bf ~~~Fig. 6.}---($U-B, B-V$)  two-color diagram for the stars in the 
inner cluster region ($r< 50\arcsec$). 
The solid lines represent the ZAMS line given by Schmidt-Kaler (1982) reddened 
by $E(B-V) =0.19$. The reddening vector with $E(B-V)$=0.20 is also shown.
Open squares denote the blue straggler star candidates. 
} 
\end{figure} 
%-------------------------------------------------- FIG 6 END 

% -------------------------------------------------- TABLE 3 START 
\begin{table*} 
\begin{center} 
{\bf Table 3.}~~Magnitudes of Blue Straggler Candidates in NGC 1193 \\ 
\begin{tabular}{llccccc} 
\hline \hline 
\noalign{\smallskip} 
X &Y & $V$ & $B-V$ & $U-B$ & $V-I$ &Comments$^\dagger$ \\ 
\noalign{\smallskip} 
\hline 
\noalign{\smallskip} 
380.052 & 505.504 & 17.179 & 0.622 & 0.074 & 0.886 &1 \\ 
567.285 & 496.319 & 16.768 & 0.684 & 0.160 & 0.925 &1 \\ 
598.199 & 472.508 & 16.890 & 0.370 & 0.202 & 0.574 &1 \\ 
487.763 & 328.471 & 17.320 & 0.600 & 0.104 & 0.859 &1 \\ 
389.350 & 232.755 & 17.303 & 0.593 & 0.085 & 0.791 &1 \\ 
718.860 & 181.164 & 16.568 & 0.654 & 0.125 & 0.884 &1 \\ 
723.588 & 366.707 & 16.009 & 0.250 & 0.145 & 0.433 &1 \\ 
801.635 & 260.945 & 15.819 & 0.366 & 0.250 & 0.497 &1 \\ 
521.599 & 236.161 & 16.509 & 0.587 & 0.124 & 0.796 &1 \\ 
571.651 & 218.347 & 15.846 & 0.373 & 0.194 & 0.466 &1 \\ 
584.537 & 185.907 & 17.067 & 0.560 & 0.101 & 0.730 &1 \\ 
552.273 & 168.404 & 15.690 & 0.486 & 0.280 & 0.655 &1 \\ 
478.316 & 146.595 & 16.307 & 0.528 & 0.202 & 0.672 &1 \\ 
633.556 & 74.153 & 16.901 & 0.413 & 0.236 & 0.558  &1 \\ 
569.258 & 60.735 & 15.463 & 0.348 & 0.246 & 0.423  &1 \\ 
589.976 & 64.463 & 16.523 & 0.344 & 0.237 & 0.510  &2 \\ 
673.547 & 191.393 & 16.906 & 0.499 & 0.136 & 0.708 &2 \\ 
466.978 & 92.772 & 17.295 & 0.509 & 0.144 & 0.703  &2 \\ 
\noalign{\smallskip} 
\hline 
\noalign{\smallskip} 
\end{tabular} 

\hspace{-1.0cm}$^{\dagger}$ : 1: Identified by Kaluzny(1988), 2: Identified by this paper
\end{center} 
\end{table*} 
%-------------------------------------------------- TABLE 2 END 

\section{PHYSICAL PARAMETERS OF NGC 1193} 

\subsection{Reddening} 

Reddening is an important basic parameter of a star clusters, 
since it can significantly affect the determination of other physical parameters. 
Several reddening values for NGC 1193 are reported. 
Kaluzny (1988) estimated the reddening to be $E(B-V)=0.12$ by isochrone fitting. 
Noriega-Mendoza \& Ruelas-Mayorga (1997) determined the reddening and metallicity values 
simultaneously from the ($V, B-V$) CMD of Kaluzny (1988), 
which are $E(B-V)=0.24$ and [Fe/H]$=-0.54$, respectively. 
If deeper $U$-band observations are made, 
the straightforward method for the reddening determination would be possible 
using the two-color diagram. 

Fig. 6 shows the ($U-B, B-V$) two-color diagram. 
The reddening value is estimated, shifting the intrinsic relation of the dwarf stars 
given by Schmidt-Kaler (1982) along the reddening line of $E(U-B)=0.72~E(B-V)$. 
In the fitting, only the stars with $U, B$, and $V$ photometric errors smaller than 
0.03 and $r<50\arcsec$ are used. 
Specially, since the dwarf line 
moves along the solid line in the diagram, the most sensitive regions to the 
change of the reddening value are $0.2 <(B-V)_0< 0.6$, which means A-F stars. 
So, we used the stars within the blue straggler region. 
Finally, $E(B-V) =0.19 \pm 0.04$ is derived from the fitting. 

%-------------------------------------------------- FIG 7 START 
\begin{figure}%[p] 
\vskip 5mm 
\epsfxsize=7cm 
\centerline{\epsffile{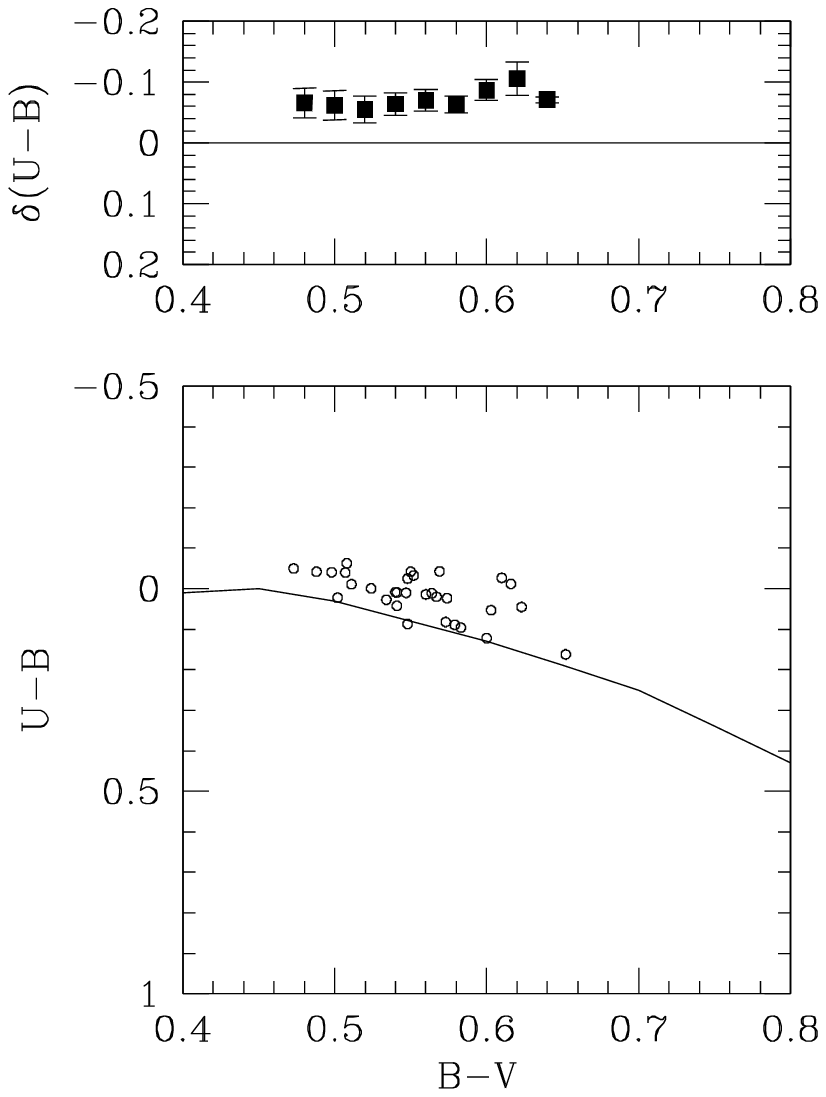}} 
{\small {\bf ~~~Fig. 7.}---Determination of metallicity from the UV excess. 
Only the stars with magnitude errors $< 0.04$ mag, $V<18.5$ mag and 
$r<50\arcsec$ are used. 
The solid line of lower panel means the Hyades fiducial line given by Sandage (1969). 
The upper panel is for the residuals between Hyades fiducial line and 
the stars of NGC 1193. 
} 
\end{figure} 
%-------------------------------------------------- FIG 7 END 

\subsection{Metallicity} 

Three values reported in the literature for the metallicity of NGC 1193: 
using photometric data, Kaluzny (1988) and Noriega-Mendoza \& Ruelas-Mayorga (1997) 
estimated it to be $-0.29$ and $-0.54$, respectively. 
Friel \& Janes (1993) determined [Fe/H]=$-0.50$ for NGC 1193 
using the spectroscopic data of four stars, 
which value is less secure as they pointed out already. 

%-------------------------------------------------- FIG 8 START 
\begin{figure}
\vskip 5mm
\epsfxsize=7cm
\centerline{\epsffile{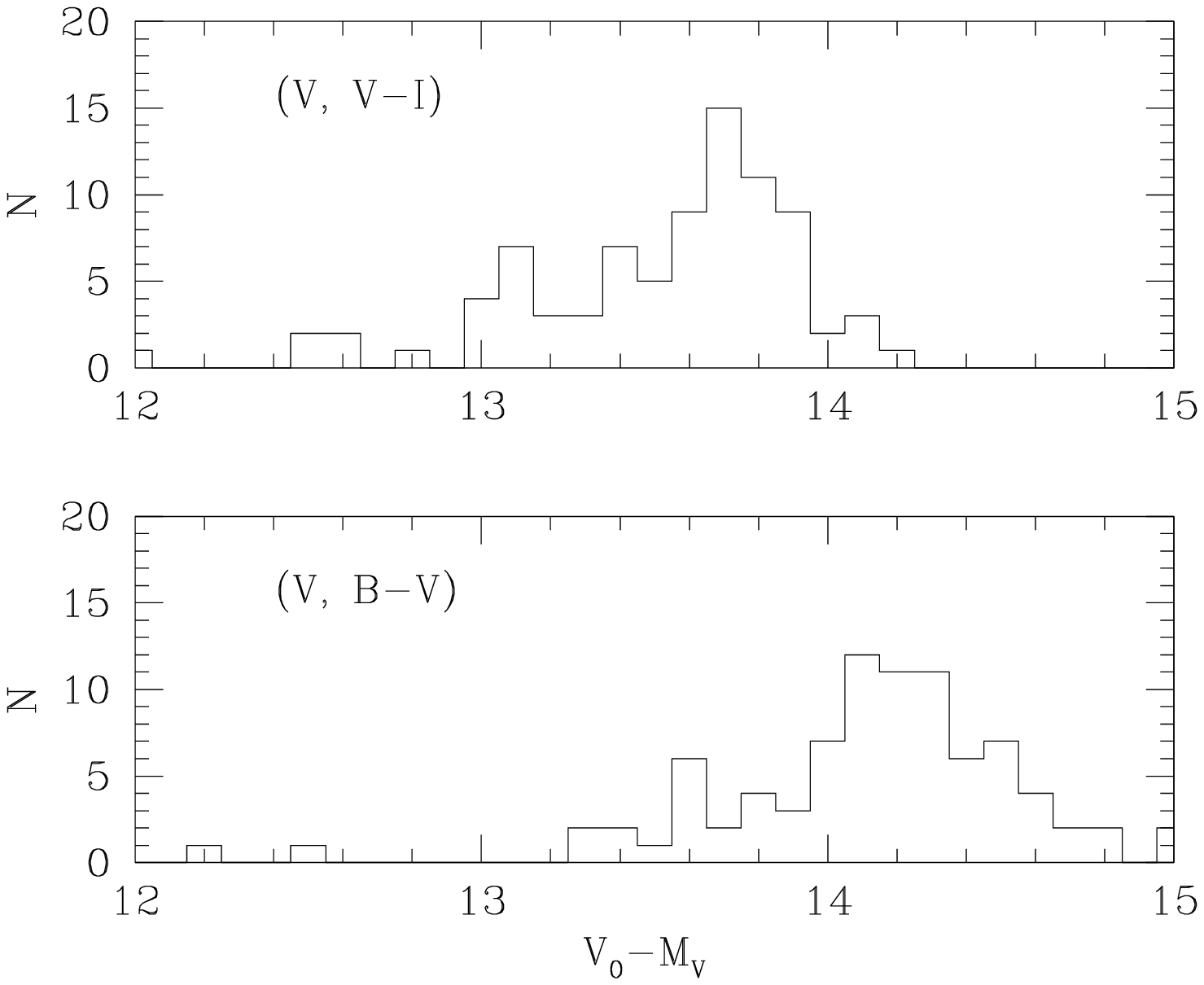}}
{\small {\bf ~~~Fig. 8.}---Distribution of distance moduli of NGC 1193 with a bin
size of 0.1 mag, in ($V,V-I$) and ($V,B-V$) CMDs.
The stars between $18 < V < 20$ and $r <50\arcsec$ are used.
The most probable value of ($m-M$)$_0$ from the ($V,V-I$) and ($V,B-V$) CMDs are
13.7 and 14.1, respectively.
}
\end{figure}
%-------------------------------------------------- FIG 8 END

The metallicity of an open cluster can be estimated 
from the difference of the distance moduli measured in 
($V, V-I$) CMD and ($V, B-V$) CMD, because the distance modulus depends on metallicity. 
Pinsonneault et al. (2004) illustrated the sensitivity of the MS 
luminosity to metallicity in three planes of ($V, B-V$), ($V, V-I$), and ($V, V-K$). 
Among these three planes, 
the metallicity sensitivity to the distance modulus is less steep at $V-I$. 
Pinsonneault et al. (2004) also constructed empirically calibrated Hyades isochrone. 
By using these characteristics, we can approximately estimate the metallicity 
of NGC 1193. 
The difference of the distance moduli measured in ($V, V-I$) and ($V, B-V$) CMDs 
after fitting their empirically calibrated isochrones to the two CMDs is 0.4 mag, 
which means the metallicity of NGC 1193 [Fe/H]$\sim -0.4$. 

More directly, using the reddening value obtained above, 
the metallicity of NGC 1193 is estimated 
using the ultraviolet (UV) excess in the ($U-B, B-V$) diagram. 
In Fig. 7, the Hyades fiducial line given by Sandage (1969) and 
the dwarf stars with $V < 18.5$ mag and $U, B, V$ photometric 
errors smaller than 0.04 mag are plotted. 
The UV excess is measured to be $\delta(U-B)_{B-V=0.6} = - 0.087 \pm 0.017$. 
The metallicity from the UV excess can be calibrated by Cameron (1985)'s relation, 
[Fe/H]$=0.08 - 4.93 \delta - 13.51 \delta^2$. 
The metallicity of NGC 1193 is thus obtained to be [Fe/H]= $-0.45 \pm 0.12$. 

%-------------------------------------------------- FIG 9 START 
\begin{figure*}%[p] 
\vskip 5mm
\epsfysize=8.9cm
\centerline{\epsffile{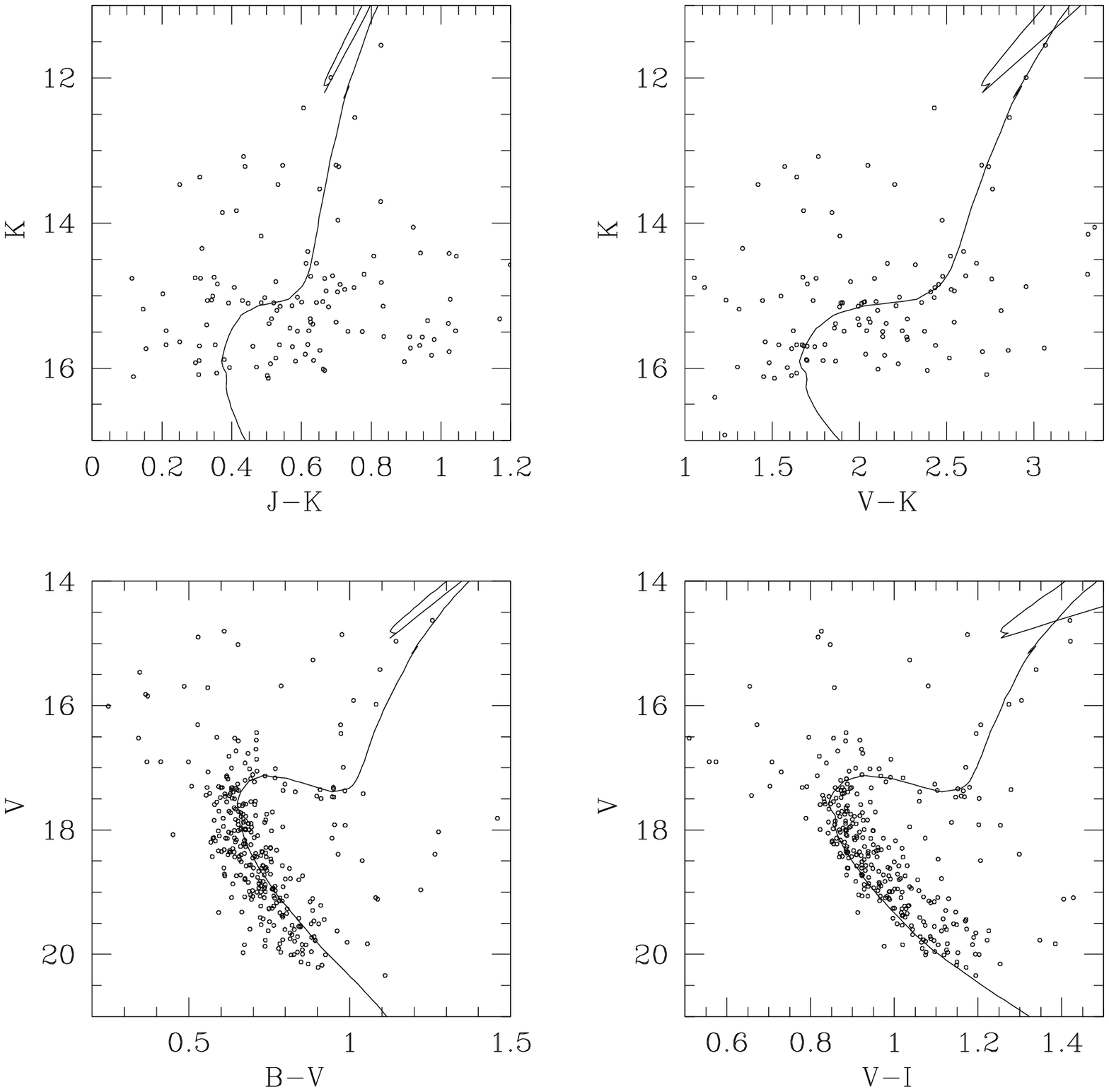}}
{\small {\bf ~~~Fig. 9.}---Padova isochrone fittings for NGC 1193
for various CMDs.  The solid lines represent the Padova isochrones for
[Fe/H]$= -0.45$ and log $t = 9.7$.
}
\end{figure*}
%-------------------------------------------------- FIG 9 END 

% -------------------------------------------------- TABLE 4 START 
\begin{table*}
\begin{center}
{\bf Table 4.}~~Basic Information of NGC 1193 \\
\vskip 3mm
{\small
\setlength{\tabcolsep}{1.2mm}
\begin{tabular}{lccl} \hline\hline
Parameter & Information & Reference \\
\hline
R.A. (J2000) & 03$^h$ 05$^m$ 56$^s$ & SIMBAD \\ %$^{\rm a}$ \\ 
Decl. (J2000) & $+44$\arcdeg~ 22\arcmin~ 59\arcsec & SIMBAD \\
$l, b$ & 146.\arcdeg81, $-12.\arcdeg16$ & SIMBAD \\
Reddening, $E(B-V)$ & $0.19 \pm 0.04$ mag & This study \\
Distance modulus, $(m-M)_0$ & $13.3\pm 0.15$ mag & This study \\
Distance, d & $4.6 \pm 0.3$ kpc & This study \\
Metallicity, [Fe/H] & $-0.45 \pm 0.12$ & This study \\
Age, $t$ & $5.0 \pm 1.2$ Gyr (${\rm log}~t=9.7 \pm 0.1$) & This study \\
\hline
\end{tabular}
} %\small 
\end{center}
\end{table*}
%-------------------------------------------------- TABLE 4 END 

\subsection{Distance Modulus} 

The distance modulus of NGC 1193 can be measured by using the ZAMS fitting. 
In this case, a priori known metallicity of the cluster is needed 
since the distance modulus is systematically shifted by the metallicity of the cluster. 
Pinsonneault et al. (2004) presented empirically calibrated isochrone of Hyades cluster 
([Fe/H]$= +0.13$) for the ZAMS fitting. 
They also gave the amount of correction 
($\Delta M_V$) as a function of [Fe/H] in several CMD planes 
in the computation of the cluster distances. 
Specially, the $M_V$ at fixed $B-V$ has an average slope 
of 1.42 mag (0.72 mag in $V-I$ CMD) per dex in metallicity. 

Fig. 8 shows the fitting results. In ($V, B-V$) CMD, 
the most probable distance modulus is $14.10 \pm 0.13$ and 
in ($V, V-I$) CMD, the value is $13.72 \pm 0.15$. 
Adopting Pinsonneault et al. (2004)'s metallicity sensitivity of isochrone 
mentioned above and considering [Fe/H]$= -0.45$ of NGC 1193 obtained above, 
the distance modulus of NGC 1193 converge into $13.32 \pm 0.15$. 

\subsection{Age and Isochrone Fitting} 

The age of a star cluster can be determined by comparing the observed CMDs 
(especially, $BVI$ passbands) with theoretical isochrones. 
Fig. 9 shows the best matched Padova theoretical isochrone (Bertelli et al. 1994) 
using the physical parameters obtained above ($E(B-V)=0.19$, [Fe/H]$= -0.45$), 
which gives the cluster age to be log $t = 9.7 \pm 0.1$. 
For the near-IR magnitudes, we sampled the $JHK_s$ data from the 2MASS 
photometry database (available at http:// irsa.ipac.caltech.edu/applications/Gator/) 
and matched them with our $BVI$ magnitude set through coordinate transformation.

In most cases of CMDs in Fig. 9, the isochrone follows the stellar sequence very well. 
The 2MASS photometry is not deep enough to clearly show the MS turn-off point 
in the near-IR bands. 

\section{SUMMARY} 

The goal of this paper is to present the $UBVI$ photometric data 
and derive the basic physical parameters of the old open cluster NGC 1193. 
The CMDs of NGC 1193 show the characteristics of the old open clusters, such as 
the sub giant branch and red giant branch.  The blue straggler candidates 
inside of $r=50\arcsec$ from the center is identified and their magnitudes are also 
measured. 

Using color-color diagrams, the reddening value is determined to be 
$E(B-V) =0.19 \pm 0.04$, which is somewhat larger 
than the Kaluzny (1988)'s value 0.12. 
The metallicity of NGC 1193 determined by using the UV excess method 
is [Fe/H]$= -0.45 \pm 0.12$, which is in the middle of the two previously values 
in the literature, $-0.50$ (spectroscopic data of four stars, Friel \& Janes 1993), 
$-0.29$ ($BV$ isochrone fitting, Kaluzny 1998). 
Our derived distance modulus is $(m-M)_0 =13.3 \pm 0.15$. 
This value is well matched with the Kaluzny's value of 13.4 within its error. 
Finally, from the Padova isochrone fitting to the observed CMDs, 
the age is estimated to be log $t =9.7 \pm 0.1$. 
The summary of the fundamental parameters for NGC 1193 is presented in 
Table 4. 

\vspace{4mm} 

We would like to thank the staff of Observatorio Astron\'omico Nacional 
in the Sierra San Pedro M\'atir (OAN-SPM) in Mexico for the use of observing 
facilities and their warm support during our observations. 
This research has made use of the SIMBAD database, operated by CDS, Strasbourg, France. 
This research has also made use of the WEBDA database, operated at the Institute 
for Astronomy of the University of Vienna.

\end{document}